\documentclass[fleqn,10pt]{wlscirep}
\usepackage[T1]{fontenc}
\usepackage[sort&compress]{natbib}
\setcitestyle{super,comma,sort&compress,open={},close={}}
\usepackage{longtable,array}
\usepackage{tcolorbox}
\tcbuselibrary{breakable,skins}
\usepackage{placeins}
\usepackage{pdfpages}

\title{A latent dimension of Condorcet's jury theorem for multiple AI advisers}
\author[1]{Kazutoshi Sasahara}
\author[1,2]{Aoi Naito}
\author[1]{Ryo Fujie}
\affil[1]{Institute of Science Tokyo, Tokyo, Japan}
\affil[2]{Carnegie Mellon University, Pittsburgh, PA, USA}
\affil[ ]{Contact: sasahara.k.6efa@m.isct.ac.jp}

\begin{abstract}
When the same question is asked of multiple AI advisers, as in self-consistency and LLM-as-a-judge panels, Condorcet's jury theorem predicts that adding independent, competent advisers makes the majority more reliable. The theorem, however, has a latent dimension when viewed from the user's vantage: adding advisers also makes disagreement more visible. A binomial model reveals that this ``visible dissent'' becomes nearly inevitable as the number of advisers grows, and that reliability and disagreement approach certainty at rates that cross at an adviser accuracy of 4/5 (0.8); below it, visible dissent eventually becomes more likely than a correct majority. Even ideal panels of independent and competent advisers can be correct in aggregate but appear divided; such disagreement does not by itself indicate aggregation failure. The way advisers split also provides a common basis for predictive multiplicity, reconciliation load, and reliance miscalibration. These results separate aggregation from disclosure and turn the latter into testable questions about how disagreement should be presented and interpreted.
\end{abstract}

\begin{document}
\flushbottom
\maketitle
\thispagestyle{empty}

\section*{Introduction}\label{introduction}

To decide what to believe, people rely on experts, institutions, and other sources of guidance, which serve as ``trust anchors''.
Yet that role is shifting to AI advisers,\cite{logg2019algorithm} especially for questions that people cannot reliably assess on their own.
No AI model, however, is perfectly accurate, and its answers are often claims that people cannot directly verify.\cite{augenstein2024factuality,suzgun2025language}
In such cases, people consult a second model and then a third, or query the same model repeatedly, in effect convening a panel of AI advisers.
For example, ask several diagnostic systems about one medical scan, and one may answer ``nothing concerning'', while another raises the possibility of cancer.\cite{vanwinkel2025ai}
Although a larger panel can yield a more reliable majority, it can also make disagreements harder for the user to ignore.

Asking more is Condorcet's jury theorem at work.\cite{condorcet1785essai}
In its modern form,\cite{grofman1983thirteen,nitzan1985collective,list2001epistemic,goodin2018epistemic} if independent advisers are each correct with probability greater than chance, the aggregate reliability ($P_{\text{agg}}$) approaches certainty as the panel grows (Supplementary Note~1).
The same logic underlies the ``wisdom of crowds''\cite{galton1907vox,surowiecki2004wisdom} and, two centuries on, systems that aggregate multiple AI advisers.
Examples include self-consistency, which relies on a majority vote over repeated samples from a single model;\cite{wang2023selfconsistency}
LLM-as-a-judge panels, in which multiple models evaluate a candidate response;\cite{zheng2023judging,verga2024replacing} and multi-model aggregation.\cite{du2024improving,chen2024reconcile}

\begin{figure}[t]
\centering
\includegraphics[width=0.8\linewidth]{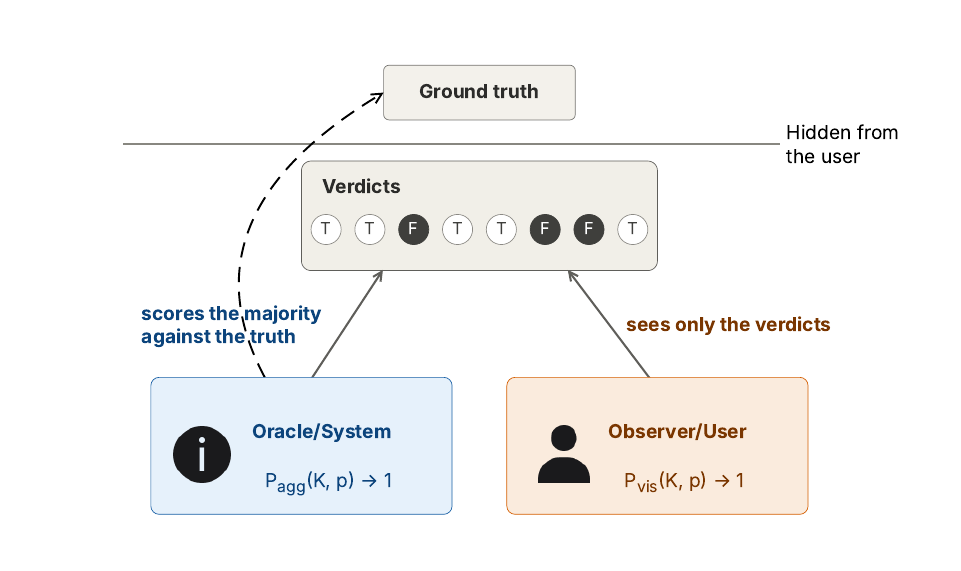}
\caption{\textbf{Two observables from the same verdicts.} The same verdicts are read from two vantages.
The system (left), an oracle with the ground truth, scores the majority against it.
The user (right) sees only the verdicts.
As the panel grows, aggregate reliability approaches certainty ($P_{\text{agg}}\to1$), while the user increasingly sees a split ($P_{\text{vis}}\to1$).}\label{fig:observables}
\end{figure}

The theorem, however, is usually read from the system's vantage (Fig.~\ref{fig:observables}, left).
It scores the majority against the ground truth, collapsing the panel to a single bit (the majority position) and discarding the vote distribution the user observes.
That distribution is a second observable.
Receiving multiple verdicts therefore introduces a form of uncertainty absent from a single verdict. 
This is one form of what we call the ``multi-trust-anchor problem'', the new uncertainties that arise once trust is distributed across many anchors.
Adding advisers makes disagreement more frequent as the majority becomes more reliable, raising a design question: how should systems help users interpret that disagreement?

Here we revisit the theorem from the user's rather than the system's vantage.
Unlike the system, the user does not observe the ground truth at the time of consultation and sees only the advisers' verdicts (Fig.~\ref{fig:observables}, right).
This epistemic shift makes another quantity relevant: the probability of ``visible dissent'' ($P_{\text{vis}}$), or how often the advisers appear to split.
As the panel grows, both aggregate reliability and visible dissent approach one, but at different rates.
The two convergence rates cross at a particular adviser accuracy.
This crossing provides a quantitative reference for studying how systems should present multiple advisers' verdicts.

To show that the multi-trust-anchor problem arises in its most basic form, even under conditions favourable to aggregation, we study this comparison in the standard binomial setting of Condorcet's jury theorem, which we use as our base model.
Each question has two possible verdicts, with one of them correct, so the analysis applies to questions with a ground truth.
This still covers many high-stakes settings, such as medical screening and fact-checking, where the decision is effectively binary.
We consider $K$ independent advisers, each answering a given question correctly with the same probability $p$, $1/2<p<1$, and otherwise free of incompetence and bias.
The equal-accuracy assumption here is made for clarity and is not required by the theorem.
Under these assumptions, disagreement need not be a sign of failure: it can arise even among ideal advisers.

\section*{A latent dimension: what the user observes}
\label{a-latent-dimension-what-the-user-observes}

\begin{figure}[t]
\centering
\includegraphics[width=0.9\linewidth]{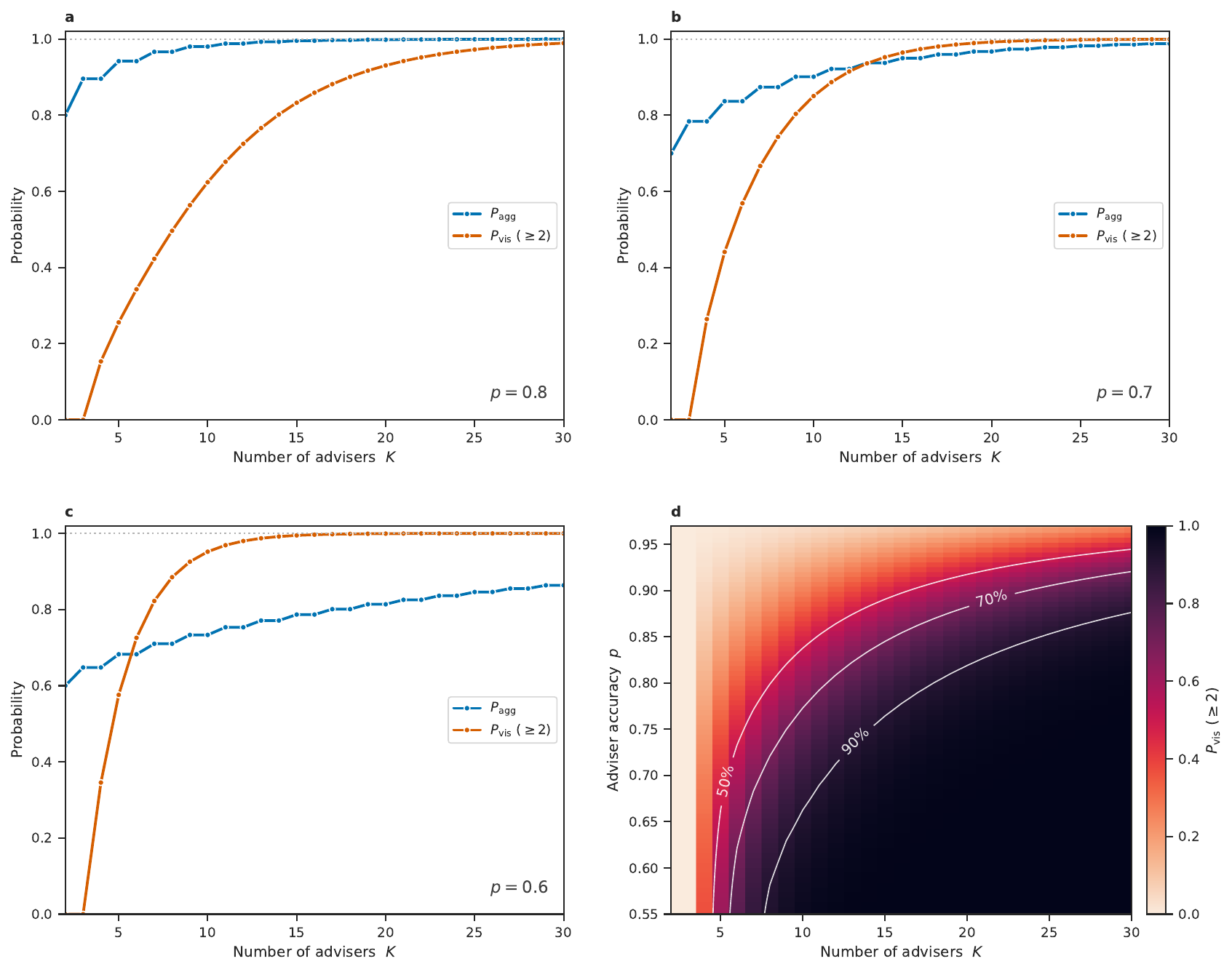}
\caption{\textbf{Disagreement is inevitable and, below $p=4/5$, eventually overtakes reliability.} These are overlapping events in the same panel, not competing outcomes. $K$ independent advisers each correct with probability $p$, $1/2<p<1$.
A split means at least two votes on each side.
\textbf{(a--c)} Aggregate reliability $P_{\text{agg}}$ (fair tie-breaking) and visible dissent $P_{\text{vis}}$ both tend to $1$ yet decouple, with the first crossover moving with accuracy.
At $p=0.8$ \textbf{(a)}, visible dissent remains less probable than a correct majority at every $K$; at $p=0.7$ \textbf{(b)}, it first overtakes at $K=14$; and at $p=0.6$ \textbf{(c)}, it first overtakes at $K=6$.
\textbf{(d)} $P_{\text{vis}}(K,p)$ across panel size and adviser accuracy, with contours at $50\%$, $70\%$, and $90\%$.}\label{fig:decoupling}
\end{figure}

From the user's vantage (Fig.~\ref{fig:observables}, right), the question is how often the panel looks divided.
We focus on the minimal split, with at least two votes on each side. 
Two rationales support this choice.
Psychologically, a lone dissenter and a shared minority can play different roles.
Breaking unanimity can substantially reduce majority pressure,\cite{asch1951effects,asch1956studies} and with panels of AI advisers a single dissent reduced conformity, whereas wider disagreement created confusion.\cite{tsuchiya2026more}
Consistent minorities can exert social influence,\cite{moscovici1969influence} and that influence depends in part on how many sources share the position.\cite{latane1981social}
We therefore use a split with at least two advisers on each side to operationalize competing positions that the user may need to reconcile, rather than a lone dissenting vote.
Mathematically, this is the smallest split that cannot be attributed solely to the error of a single adviser. Even an ideal panel will often include a lone dissenter, since each adviser errs at a rate of $q$ ($=1-p$).
This threshold is an operational definition of visible dissent, not a claim that lone dissent never matters.

We use this minimal split throughout. In the generic form, the count of two is replaced by a fixed fraction $r$ of the panel on each side (Supplementary Notes~3 and~4).
The complement of this event consists of the four near-unanimous configurations: zero, one, all but one, or all advisers on one side.
Under this definition, the probability of visible dissent is expressed in closed form as
\[
P_{\text{vis}}(K,p) = 1 - q^{K-1}(Kp + q) - p^{K-1}(Kq + p),
\]
which approaches $1$ for all $p < 1$. Equivalently, near-unanimous panels become rare.
Both facts follow from the binomial distribution alone (Supplementary Note~2). 
The closed form of visible dissent is symmetric in $p$ and $q$: a split indicates that the panel disagrees, but the event itself is side-blind.

Which of aggregate reliability and visible dissent is larger, and for how long, depends on adviser accuracy.
At $p = 0.8$, a split never overtakes a correct majority (Fig.~\ref{fig:decoupling}a), whereas at lower accuracies it does (Fig.~\ref{fig:decoupling}b,c).
Even panels with near-perfect accuracy routinely exhibit splits.
This structural decoupling across panel sizes and accuracies is captured by $P_{\text{vis}}$ (Fig.~\ref{fig:decoupling}d).
At $p = 0.9$ and $K = 20$, the majority is almost certain to be correct (above 99.999\%), yet the user sees a split about $61\%$ of the time.
Agreement can still be common on easy questions, where the per-item adviser accuracy $p$ is close to one.
For any fixed imperfect accuracy in the base model, splits approach certainty as the panel grows, although higher accuracy delays this increase.
A human observer may find such disagreement confusing, even when the majority verdict is correct.
Beyond whether a panel splits, users can observe how large the minority is.
The dissenting fraction $m$ is the proportion supporting the minority answer.
As the panel grows, $m$ approaches $q$, even as the majority becomes increasingly likely to be correct (Supplementary Note~3).

\section*{When visible dissent outpaces reliability}
\label{when-visible-dissent}

\begin{figure}[t]
\centering
\includegraphics[width=\linewidth]{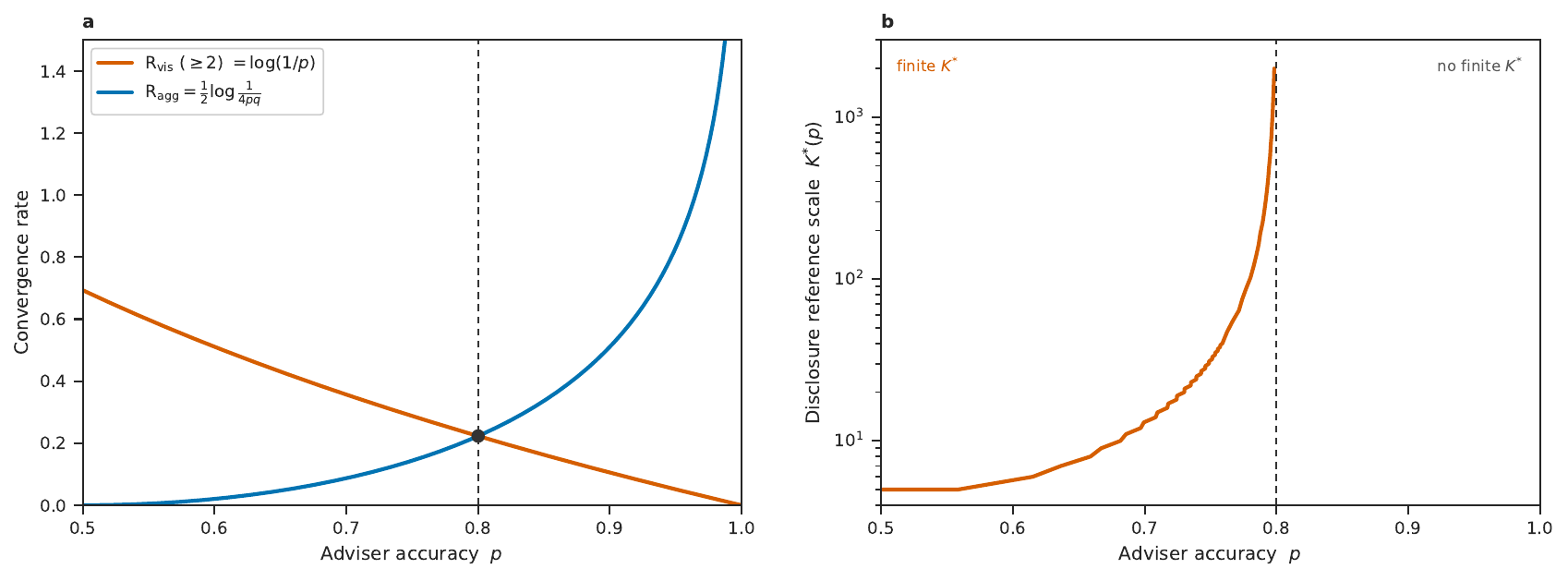}
\caption{\textbf{The two convergence rates cross at $p=4/5$.} \textbf{(a)} Convergence rates of aggregate reliability and visible dissent. A finite disclosure reference scale $K^{*}$ exists for $p<4/5$ and not for $p\ge4/5$.
\textbf{(b)} $K^{*}(p)$, the panel size at which the first crossover occurs (a split becomes at least as likely as a correct majority), diverging as $p$ approaches $4/5$ from below (log scale).
$K^{*}=6,14,32$ at $p=0.6,0.7,0.75$ under fair tie-breaking. The values shift under a strict-majority convention (Supplementary Note~5).
The vertical dashed lines mark $p=4/5$. Benchmark-average accuracy is not equivalent to the per-item $p$ shown here.}\label{fig:rates}
\end{figure}

Which approaches certainty faster, aggregate reliability or visible dissent?
Both approach it exponentially fast as $K$ grows, and the two convergence rates can be compared directly.
The former is the classical error exponent of majority voting, and the latter the standard large-deviation rate at which near-unanimous panels become rare\cite{dembo1998large} (Supplementary Note~4).
This comparison shows that the two rates cross at a single adviser accuracy $p=4/5$ (0.8, or odds of four to one).
Above this line, aggregate reliability converges faster and stays ahead, whereas below it the order reverses and visible dissent converges faster (Fig.~\ref{fig:rates}a).
Figure~\ref{fig:decoupling}a sits exactly on this line ($p=0.8$), where reliability keeps its lead at every finite panel size. The lower-accuracy panels of Fig.~\ref{fig:decoupling}b,c fall below it ($p=0.7$ and $p=0.6$, respectively).
This crossing is robust: changing the fixed-count split threshold shifts the finite crossover size but not the $4/5$ boundary, while more general fraction thresholds place the boundary below $4/5$ and approach it as the fraction tends to zero (Supplementary Fig.~1).
Giving $1-P_{\text{vis}}$ and $1-P_{\text{agg}}$ different fixed weights can likewise shift the first crossover but leaves the $4/5$ boundary unchanged.
Under the base model, no split definition that excludes unanimity can approach certainty at a faster exponential rate than aggregate reliability at or above $p=4/5$ (Supplementary Note~4).

Below the $4/5$ line the probability ordering of the two quantities can reverse, and two effects must be kept apart.
The first is that visible dissent rises with $K$ at every accuracy.
The second is the reversal of the probability ordering, $P_{\text{vis}}\ge P_{\text{agg}}$.
Because the two events overlap, this ordering is equivalent to comparing incorrect-majority splits with correct near-unanimous panels (Supplementary Note~4 and Supplementary Table~1).
At or above the line, a correct majority stays more likely than a split, no matter how large the panel grows.
We call the value $4/5$ the ``rate boundary'' of the base model.

The first panel size at which visible dissent becomes at least as likely as a correct majority defines the ``disclosure reference scale'' $K^{*}(p)$.
At this first crossover, panels in repeated consultations are divided at least as often as their majority is correct---and a correct majority can still come from a divided panel.
The rate boundary at $p=4/5$ determines whether such a finite size exists.
For $p \ge 4/5$, an analytic proof rules out any crossover (Supplementary Note~5).
Near the line, $K^{*}$ grows rapidly, reaching hundreds of advisers, with both probabilities extremely close to one. At $p=0.6$ and $0.7$, by contrast, $K^{*}$ is 6 and 14 (Fig.~\ref{fig:rates}b).

The design question therefore arises even before independence or competence fails: how should a system present verdicts that become more reliable in aggregate yet more likely to appear divided?
The rate boundary and $K^{*}$ provide quantitative reference points for investigating this question.
We next connect these results to prior work, human responses to disagreement, and dependence among advisers.
\FloatBarrier

\section*{Two vantages on the same disagreement}
\label{two-vantages-on-the-same-disagreement}

The disagreement itself is not new. 
In predictive multiplicity, ``ambiguity'' is the proportion of inputs on which similarly accurate models make conflicting predictions.\cite{marx2020predictive,watsondaniels2023predictive,black2022model} Earlier, Poisson and, later, Gelfand and Solomon examined how verdicts split across a panel.\cite{poisson1837recherches,gelfand1973study} 
These lines of work ask different questions about agreement and disagreement. Table~\ref{tab:landscape} organizes them around two vantages: collective performance from the system's side, and exposure to and interpretation of advisers' outputs from the user's side. 
A closely related approach conditions on the realized vote to infer competence and truth,\cite{romeijn2011learning} treating the observed vote margin as evidence rather than asking how often dissent will be seen at all.
Work on AI aggregation asks how disagreement among advisers affects collective accuracy, while studies of multi-AI advice examine how observed agreement and disagreement affect user reliance.\cite{lu2024does,tsuchiya2026more}
We connect these questions by comparing aggregate reliability with the probability of visible dissent as a panel grows, even under classical assumptions favourable to aggregation.

Two reference quantities characterize this comparison: the adviser accuracy ($p=4/5$) below which visible dissent eventually surpasses aggregate reliability, and the corresponding disclosure reference scale ($K^{*}$).
The result is not specific to AI: it applies to any collection of independent, imperfect advisers with the same accuracy, including jurors, sensors, and models. 
AI is unusual in that users can assemble such panels on demand. 
To our knowledge, this comparison of the convergence of visible dissent and aggregate reliability, together with its use as a reference for disclosure design, has not been developed in prior work.

\begin{table}[t]
\centering
\caption{\textbf{Two vantages on disagreement among advisers.}
View identifies the analytical perspective emphasized here: evaluating collective performance (Oracle/System) or interpreting the outputs available to an observer (Observer/User).
These perspectives can overlap within a study.
This work examines exposure to disagreement under the classical ideal assumptions, comparing visible-dissent probability with aggregate reliability.}\label{tab:landscape}
\footnotesize
\begin{tabular}{@{}>{\raggedright\arraybackslash}p{0.14\linewidth}>{\raggedright\arraybackslash}p{0.18\linewidth}>{\raggedright\arraybackslash}p{0.24\linewidth}>{\raggedright\arraybackslash}p{0.36\linewidth}@{}}
\toprule
\textbf{View} & \textbf{Focus} & \textbf{Question} & \textbf{Representative work} \\
\midrule
Oracle/System & Dependence, diversity and participation & What shapes the reliability of collective judgements? & Dependent software failures\cite{knight1986experimental}; algorithmic monoculture\cite{kleinberg2021algorithmic}; ensemble diversity\cite{krogh1995neural,kim2026diversity}; interpreted versus generated signals\cite{hong2009interpreted}; pooling medical judgements\cite{kurvers2016boosting}; confidence-based abstention\cite{karge2026epistemic} \\
Oracle/System and Observer/User & Vote distributions and multiplicity & How do verdicts differ across advisers and models? & Jury verdict-split models\cite{poisson1837recherches,gelfand1973study}; ensemble diversity measures\cite{kuncheva2003measures}; predictive multiplicity\cite{marx2020predictive}; positive dissensus\cite{landemore2015deliberation} \\
Observer/User & Uncertainty and inference & What can advisers' outputs tell us about uncertainty, competence, or truth? & Deep ensembles\cite{lakshminarayanan2017simple}; disagreement-based UQ\cite{jiang2026discouq}; abstention on arbitrary predictions\cite{cooper2024arbitrariness}; abstention based on vote thresholds\cite{nairkanneganti2025increasing}; posterior updating from jury size and margin\cite{romeijn2011learning}; algebraic evaluation from answer patterns\cite{corradaemmanuel2024algebraic}; consensus--accuracy laws across models\cite{liu2026jury}; unanimity as evidence of systematic failure\cite{gunn2016toogood}; correlation neglect\cite{enke2019correlation}; decisions with misleading correlated signals\cite{fehrler2026toogood} \\
Observer/User & Visible dissent under the classical ideal assumptions (\textbf{this work}) & How often will users encounter a split as aggregate reliability improves? & Visible-dissent probability of independent, equal-accuracy, better-than-chance advisers compared with aggregate reliability; rate boundary $p=4/5$; disclosure reference scale $K^{*}$ \\
\bottomrule
\end{tabular}
\end{table}

\section*{Three perspectives on the same vote split}
\label{three-perspectives-on-the-same-split}

\begin{figure}[t]
\centering
\includegraphics[width=\linewidth]{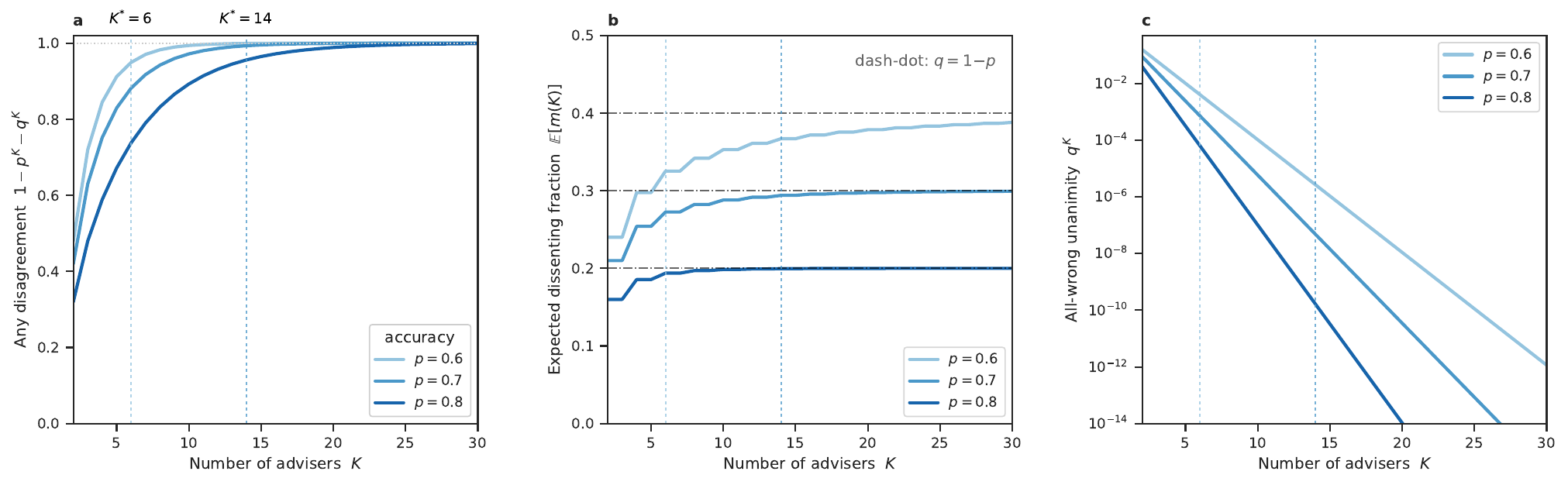}
\caption{\textbf{A common predictive substrate.} All three quantities are derived from the distribution of the same vote split $Y$ under the base model, with $q=1-p$.
\textbf{(a)} The probability of any disagreement, $1-p^{K}-q^{K}$, which predictive multiplicity reads as the rate of conflicting predictions. Unlike $P_{\text{vis}}$, it counts even a single dissenting vote.
\textbf{(b)} The expected dissenting fraction $\mathbb{E}[m(K)]$, which approaches $q$ as the panel grows and is a candidate stimulus for reconciliation load.
\textbf{(c)} The probability of all-wrong unanimity, $q^{K}$ (log scale), relevant to over-trust. From the verdicts alone, users cannot distinguish a unanimously wrong panel from a unanimously correct one.
Dotted vertical lines mark $K^{*}(p)$ for reference.
The curves are model-derived quantities, not behavioural responses. The link in (b) to reconciliation load rests on the monotone-response hypothesis (Supplementary Note~7).}
\label{fig:substrate}
\end{figure}

Why not simply ignore the split and act on the majority?
The majority remains the best prediction, but this does not make the split irrelevant.
Three separate literatures approach observed disagreement from different angles.
All three begin from the same information available to the user: the vote split (Fig.~\ref{fig:substrate}).
Predictive multiplicity---mentioned above---reads it as the probability of observing any disagreement.
Reconciliation load---the cognitive cost of reconciling conflicting advice\cite{yaniv2007using,tsuchiya2026more}---is linked here to the dissenting fraction ($m$).
Reliance miscalibration---either under-reliance or over-trust in the model panel\cite{dietvorst2015algorithm,zhang2020effect,bucinca2021to,passi2022overreliance,schemmer2023appropriate}---can arise from exposure to a split and from unanimity, which looks the same whether all advisers are correct or all are wrong (probabilities $p^{K}$ and $q^{K}$).
While each literature has measured its quantity after the fact, the base model gives each a predictive counterpart in advance. Testing these predictions defines an experimental agenda.

The predictive-multiplicity reading is exact under the base model's assumptions.
For reconciliation load and under-reliance, the base model specifies the stimulus available to the user, while the behavioural response remains outside the base model.
To fill this gap, we assume that each behavioural response does not decrease as its corresponding stimulus increases.
We call this the ``monotone-response hypothesis'' and leave its empirical test to future work (Supplementary Note~7).

Under this hypothesis, expected reconciliation load does not decrease with greater dissent.
Exposure to a split may also increase under-reliance, since many splits occur despite a correct majority and may be interpreted as evidence of uncertainty.
In the base model, the observed margin is itself evidence.
At $p=0.7$ and $K=14$, majority accuracy is $0.938$ before observing the vote, but $0.845$ conditional on an 8--6 split, a margin of two votes  (Supplementary Note~7).
Lowering trust after a narrow split can therefore be a calibrated response rather than under-reliance.

Over-trust has a different source. 
An all-wrong panel occurs with probability $q^{K}$ under independence and presents the same unanimous vote as an all-correct panel.
Following a unanimous panel is well founded under independence, where it is correct with probability $1/(1+(q/p)^{K})$. Over-trust can arise when reliance on unanimity exceeds what the evidence warrants, for example when correlated errors inflate all-wrong unanimity above the independence baseline but users treat the signals as independent.

The base model also specifies how each stimulus varies with panel size and accuracy, providing predictions that experiments can test. 
All three considerations can therefore be placed on the same panel scale, with $K^{*}$ marking the first crossover. 
The stimuli for reconciliation load and under-reliance are already substantial at this scale, whereas the all-wrong probability $q^{K}$ has become negligible under independence.

\section*{From null model to exact sampling law}
\label{sec:null-model-machine-internal}

Pairwise disagreement among frontier models of comparable accuracy has been measured at $16$--$38\%$ on multiple-choice benchmarks,\cite{yang2026benchmark} but these rates also reflect the answer space and shared training data or lineage,\cite{kuai2026how,kohli2026nine} and therefore do not map directly onto per-item accuracy.
Real AI advisers are rarely independent or equally accurate.
Even so, the base model provides an independence baseline: the distribution of vote splits an equal-competence panel would show if its members voted independently.
A real panel can fall short of this baseline for reasons unrelated to dependence, such as easy questions or unequal accuracies.
After adjusting for these factors, a remaining deviation that exceeds estimation and sampling error provides evidence of residual dependence. We call a shortfall of this kind the ``dissent deficit'' (Supplementary Fig.~2a).
Extensions of the theorem to correlated votes show that dependence alters aggregate reliability and can erode the gains that independence provides.\cite{ladha1992condorcet,boland1989majority} In the correlated panels we simulate (one-factor Gaussian copula; Supplementary Note~6), it also lowers the split rate, although this does not follow from positive pairwise correlation alone.
This independence baseline can also help calibrate methods that use disagreement as a signal of uncertainty.\cite{jiang2026discouq}

Positive dependence can also make agreement less reassuring.
Advisers that make similar errors can produce a unanimous panel that is simply wrong, and such all-wrong panels can arise more often than the $q^{K}$ predicted under independence (equal-competence case; Supplementary Fig.~2b).
Individually reliable advisers can fail together without any adviser having malfunctioned.
Related work on the ``paradox of unanimity'' shows how allowing for systematic failure can make overwhelming agreement less reassuring.\cite{gunn2016toogood}
Experiments with state-dependent signal correlation further show that both individuals and small groups struggle to use potentially misleading evidence, often treating correlated signals as independent.\cite{fehrler2026toogood}

The base model plays a different role in machine-internal sampling.
A system that generates and aggregates multiple outputs can observe how they split without making any assumption about human responses.
The observer need not be human: such a system stands at the same vantage, seeing the split but not the ground truth.
The clearest case is repeated sampling from a fixed model. 
If responses to the same item are drawn independently from a fixed output distribution under the same prompt and generation settings, and each is mapped to one of two verdicts, the induced verdicts---and therefore their correctness indicators---are i.i.d., with probability $p$ of being correct. The panel size $K$ is the number of samples drawn.
Here the closed form for $P_{\text{vis}}(K,p)$ is not a baseline but an exact law.
Changes in prompts, model versions, or serving conditions can cause the samples to no longer be i.i.d., taking the system outside the base model.
Within this exact law, sampling beyond $K^{*}$ can still improve aggregate reliability for $p<4/5$, so $K^{*}$ does not say when to stop sampling but provides a reference for comparing how the samples are reconciled or presented.

The closed form does not require the model to be better than chance.
On an item where $p$ falls below $1/2$, the samples can remain i.i.d.\ even as majority voting becomes increasingly likely to select the wrong answer.
The split cannot warn of this because its distribution is symmetric in $p$ and $q$.
If majority accuracy stops improving as more samples are drawn, the model may have reached a competence ceiling, or the samples may no longer be independent.
After adjusting for item difficulty and unequal accuracies, a lower-than-expected split rate provides evidence of residual dependence.
In this setting, the base model helps diagnose whether further sampling adds independent information and whether reconciliation may be useful.

\section*{A framework for the disclosure layer}
\label{a-rule-for-the-disclosure-layer}

\begin{tcolorbox}[float=t!,colback=blue!5!white,colframe=blue!75!black,title={Box 1 \textbar{} A disclosure framework},before upper={\setlength{\parindent}{0pt}\setlength{\parskip}{5pt}}]

\emph{Base model.}
$K$ independent advisers, each correct with probability $p$, $1/2<p<1$, on a binary question; the majority is taken with fair tie-breaking, and a split means at least two votes on each side.

\emph{What the model gives.}
For a given $p$, the model gives the probability of encountering a split, $P_{\text{vis}}(K,p)$, and the expected dissenting fraction, $\mathbb{E}[m(K)]$, which approaches $q=1-p$ as $K$ grows.
It also gives the disclosure reference scale $K^{*}(p)$: the first panel size at which a split is at least as likely as a correct majority, finite only for $p<4/5$.

\emph{Using the framework.}
Estimate $p$ with uncertainty for the setting of interest, compute $P_{\text{agg}}$ and $P_{\text{vis}}$ at the actual $K$, and record the pre-aggregation vote distribution.
Use the table to select panel sizes at which to evaluate the same presentation policies on the outcomes below, accounting for the observed vote margin.
The table indicates where to test whether their relative effects change, not which policy to choose.

\emph{Policies to compare.}
Individual verdicts; verdicts with the vote margin and estimated confidence; an openly reconciled answer with estimated confidence and access to the full panel.
Each policy explains the split expected under the base model.

\emph{Outcomes.}
Judgement accuracy, reliance calibration conditional on the observed vote, and reconciliation load.

A panel can be both correct and split; these are overlapping events.

\begin{center}
\renewcommand{\arraystretch}{1.15}
\begin{tabular}{@{}
>{\raggedright\arraybackslash}p{0.12\linewidth}
>{\raggedright\arraybackslash}p{0.40\linewidth}
>{\raggedright\arraybackslash}p{0.40\linewidth}@{}}
\toprule
 & \textbf{$P_{\text{vis}}<P_{\text{agg}}$ at the actual $K$} & \textbf{$P_{\text{vis}}\ge P_{\text{agg}}$ at the actual $K$} \\
\midrule
$p\ge4/5$
& Holds at every $K$, although splits still become common. Use this non-reversing regime as a comparison condition for the same policies.
& Does not occur. \\[4pt]
$p<4/5$
& A finite $K^{*}$ marks the first crossover. Evaluate the policies at panel sizes around $K^{*}$ that span both orderings.
& A split is at least as likely as a correct majority. Test whether the relative effects of the same policies differ between the two orderings. \\
\bottomrule
\end{tabular}
\end{center}

\emph{Conditions of use.}
If values within the uncertainty interval for $p$ yield different orderings at the actual $K$, leave the ordering unassigned.
Treat domain-level estimates of $p$ as reference values, not item-level parameters.
Allow for item difficulty and for unequal or dependent advisers.
Apply only to questions with a ground truth.
If a minority answer identifies a potentially serious risk, keep that warning visible even when presenting a reconciled answer.

\end{tcolorbox}

What should the system show to the user, and when?
Even in the best case, where every adviser is competent and independent, two numbers need attention.
One is the dissenting fraction expected by default, $m \to q$.
The other is the reference scale $K^{*}$.
Adding advisers and showing their verdicts are distinct design decisions.
Both are handled at the disclosure layer, in how the system combines and presents its outputs (Box~1).

Users convene these panels, but only the system that presents the verdicts can act on the split.
The application of the framework depends on who receives the verdicts.
When a machine collects them and takes a majority vote, as in self-consistency, the base model yields $K^{*}(p)$ exactly.
Pipelines that do not vote, such as best-of-$N$ selection or LLM-as-a-judge scoring, fall under the framework only after their outputs are converted into verdicts and put to a majority vote.

For human-facing use, a domain-level accuracy estimate can be inserted into the base model to obtain a reference scale.
A domain average, however, is not equivalent to a mixture of easy and hard questions, because both probabilities are nonlinear in $p$.
At $K=14$, for example, an equal mixture of items with $p=0.51$ and $p=0.89$ (mean $0.7$) reverses the ordering that holds when every item has $p=0.7$ (Supplementary Note~6).
This $K^{*}$ is therefore a domain-level guide rather than an item-level trigger.
Using $K^{*}$ in practice depends on the estimate of $p$ (Box~1).
The realized split carries item-level information at run time, although it does not identify item difficulty.

The comparison of $P_{\text{vis}}$ with $P_{\text{agg}}$ highlights a potential gap between obtaining a reliable answer and knowing how to interpret the advice received.
Reliable aggregation does not remove the user's task of interpreting disagreement.
The split event itself is side-blind.
Near-agreement, by contrast, points to a side, since a near-unanimous panel is correct with probability approaching one (Supplementary Note~2).

A well-aligned model should tell the user what its verdicts show, not a correct answer it does not have.\cite{steyvers2025what}
The system should therefore explain the disagreement expected under the base model while retaining the information carried by the observed vote margin.
Presenting disagreement without that context may make even a reliable majority answer look doubtful, leaving users to resolve uncertainty that aggregation has already reduced.
Studies with two advisers point the same way.
For a second opinion to improve appropriate reliance on the AI, users should not see too many disagreements on questions that the AI has answered correctly.\cite{lu2024does}

\FloatBarrier

\section*{Discussion}
\label{discussion}
The multi-trust-anchor problem begins even in Condorcet's ideal world.
Adding independent, competent advisers makes the majority more reliable, but also makes disagreement more frequent.
Such disagreement does not, by itself, indicate a failure of aggregation.
The gain in accuracy therefore does not settle the design of the consultation: how the verdicts are presented and interpreted remains a separate question.
The $4/5$ rate boundary and $K^{*}$ make this question concrete.

The same model also provides an independence baseline for studying real adviser panels.
Three questions follow for empirical work: whether the presentation policies in Box~1 differ in their effects on judgement accuracy, reliance calibration, and reconciliation load around $K^{*}$ and across the two probability orderings; whether behavioural responses track the modelled stimuli, as the monotone-response hypothesis predicts; and how far real panels depart from the independence baseline.
A practical first step is to record and evaluate the pre-aggregation vote distribution and split rate alongside majority accuracy.
Box~1 uses these quantities to frame comparisons of presentation policies, distinguishing model results from hypotheses to test. 
The distinction between aggregation and presentation remains relevant even if policy effectiveness shows no particular change around $K^{*}$.
Model accuracy is only part of the design problem. Applying the framework also depends on how closely real adviser panels match the assumptions of the base model.

The assumptions of the base model point to the next steps, starting with independence. 
Social influence and collective exposure diminish it.\cite{lorenz2011how,muchnik2013social}
In the correlated panels we simulate, dependence suppresses visible dissent below the independence baseline (Supplementary Note~6). 
After adjusting for item difficulty and unequal accuracies, a dissent deficit provides evidence of residual dependence, but its magnitude must be calibrated before it can be used to infer how strong that dependence is. 
The competence assumption may also fail. 
When advisers are worse than chance, the majority becomes more confidently wrong as the panel grows. 
Where per-adviser accuracies are available, heterogeneity under independence is absorbed exactly by the Poisson-binomial baseline (Supplementary Note~4). 
A further extension is the connection between visible dissent and human behaviour. Only the human-facing use of the disclosure framework depends on that connection. 
The binary setting is less restrictive than it seems, since many systems turn open-ended responses into true-or-false decisions. 

A harder case for the base model arises when advisers do not answer in parallel. 
When agents see and revise one another's answers, interaction itself can shape agreement.
Human dyad experiments highlight the role of communicating confidence in joint perceptual decisions.\cite{bahrami2010optimally}
LLM experiments also show that simulated peer consensus can induce incorrect answers to questions answered correctly in isolation, and a peer arguing for the correct answer reduces these errors.\cite{kumarappan2026sycophancy}
Recent work on LLM naming games shows how mutual learning from sampled outputs can generate consensus without external evidence.\cite{tanaka2026lottery}
Extending the present framework to such settings requires tracking how interaction changes both the verdict distribution and its interpretation.

The framework does not apply equally to all questions. 
Reconciliation is appropriate when a question has a ground truth. 
For instance, whether two drugs interact has a ground truth, whereas questions of policy fairness are typically normative rather than objectively settled. 
When a system handles millions of queries, treating a contested issue as settled can hide legitimate disagreement from users. 
Deciding whether a question falls within scope is itself a governance decision and should not be left to an AI system alone.

Visible dissent is only the first layer of the multi-trust-anchor problem. 
Once the assumptions of the base model are relaxed, even agreement becomes harder to interpret: users can see that advisers agree, but not whether they reached the same answer independently or failed together. 
Comparing the observed split rate with the independence baseline offers a first diagnostic, but more complex systems introduce further layers of trust. 
One AI may be asked to judge which other AI should be trusted, and in agentic systems several components may combine outputs before anything reaches a human user. 
At each layer, the system must decide which outputs to retain, how to reconcile them, and what to disclose. 
As trust becomes distributed across multiple AI advisers, the problem shifts from the reliability of individual outputs to how agreement and disagreement are combined, interpreted, and disclosed.

\begin{center}\rule{0.5\linewidth}{0.5pt}\end{center}

\section*{Declarations}\label{declarations}

\noindent\textbf{Author contributions.} K.S. conceived and designed the study, developed the theory and the analysis, and wrote the manuscript. A.N. contributed to the design of the study by connecting the framework to research on human--AI reliance, and to the interpretation of the user-side quantities. R.F. contributed to the design of the theoretical analysis and checked the derivations in the Supplementary Information. All authors reviewed and approved the manuscript.

\noindent\textbf{Competing interests.} The authors declare no competing interests.

\noindent\textbf{Data availability.} No empirical datasets were generated or analysed in this study. The numerical analyses reported here derive from the closed forms and from synthetic simulations described in the Supplementary Information.

\noindent\textbf{Code availability.} The analytical results follow from the derivations in the Supplementary Information; the scripts that reproduce the numerical calculations and simulation figures are available at \url{https://github.com/Sasahara-Lab/mta0_code}.

\noindent\textbf{Use of AI tools.} Generative-AI assistance was used for language editing.
All technical content is the authors' own.

\bibliographystyle{naturemag}
\bibliography{refs}
\clearpage


\section*{Supplementary material}

# Supplementary Information for "A latent dimension of Condorcet's jury theorem for multiple AI advisers"

Kazutoshi Sasahara[1], Aoi Naito[1,2], Ryo Fujie[1]

[1]Institute of Science Tokyo, Tokyo, Japan [2]Carnegie Mellon University, Pittsburgh, PA, USA

Supplementary Notes 1–7 provide the derivations behind the main text: the two closed forms and their domains (Notes 1 and 2), the dissenting fraction (Note 3), the convergence rates and the boundary $p = 4/5$ (Note 4), the disclosure reference scale $K^*$ (Note 5), the independence baseline and its corrections (Note 6), and the three user-side readings of the split (Note 7).

Unless noted otherwise, the base model is homogeneous: $K$ advisers answer a binary question independently, and each is correct with probability $p$, $1/2 < p < 1$ ($q = 1 - p$). Let $Y \sim \mathrm{Bin}(K, p)$ count the correct verdicts. The user does not observe which side is correct, but the user-side quantities $P_{\mathrm{vis}}$, $m$, and the apparent-unanimity probability $U$ depend only on how the verdicts split. Only $P_{\mathrm{agg}}$ (and therefore $K^*$) uses the ground-truth label.

## Supplementary Note 1 | Aggregate reliability $P_{\mathrm{agg}}$

Under fair tie-breaking, $P_{\mathrm{agg}} = \Pr[Y > K/2] + \frac{1}{2}\Pr[Y = K/2]$ is non-decreasing in $K$[1] and approaches 1 as $K \to \infty$.[2,3] The latter follows from the strong law of large numbers (SLLN), because $Y/K \to p > 1/2$. Fair tie-breaking removes the even–odd sawtooth: $P_{\mathrm{agg}}$ is constant on each consecutive odd–even pair, $P_{\mathrm{agg}}(2j) = P_{\mathrm{agg}}(2j-1)$. Together with the increasing $P_{\mathrm{vis}}$, this explains why the ordering can depend on parity near $K^*$ (Supplementary Note 5).

## Supplementary Note 2 | Visible dissent $P_{\mathrm{vis}}$

Visible dissent is absent exactly in the four near-unanimous configurations $\{Y \le 1\} \cup \{Y \ge K-1\}$, with $P(Y \le 1) = q^{K-1}(q + Kp)$ and $P(Y \ge K-1) = p^{K-1}(Kq + p)$. Subtracting their probabilities from one gives

$$P_{\mathrm{vis}}(K,p) = 1 - q^{K-1}(Kp+q) - p^{K-1}(Kq+p), \qquad q = 1-p,$$

using only the binomial distribution. The four configurations are distinct for $K \ge 3$, and a two-per-side split exists only for $K \ge 4$. We therefore use the closed form for $K \ge 4$ and set $P_{\mathrm{vis}} = 0$ for $K \le 3$. At $K = 3$ the formula itself returns 0; at $K = 2$ the four configurations overlap, so the formula does not apply. The closed form is symmetric, $P_{\mathrm{vis}}(K,p) = P_{\mathrm{vis}}(K,q)$, and $P_{\mathrm{vis}} \to 1$ as $K \to \infty$ for every $p < 1$, because each term in the complement is a polynomial factor times $\rho^{K-1}$ with $\rho \in \{p,q\} \subset (0,1)$. It is also strictly increasing in $K$: for $K \ge 3$ and $0 < p < 1$, the closed form gives $\left[1 - P_{\mathrm{vis}}(K)\right] - \left[1 - P_{\mathrm{vis}}(K+1)\right] = K\left(p^2 q^{K-1} + q^2 p^{K-1}\right) > 0$, while $P_{\mathrm{vis}} = 0$ for $K \le 3$ by definition.

This symmetry is structural, not merely algebraic. If $Y \sim \mathrm{Bin}(K,p)$, then $K - Y \sim \mathrm{Bin}(K,q)$. Swapping $p$ and $q$ simply relabels which side is correct ($Y \leftrightarrow K - Y$). Visible dissent, $\{2 \le Y \le K-2\}$, is unchanged by this relabelling, so $P_{\mathrm{vis}}(K,q) = P_{\mathrm{vis}}(K,p)$. By contrast, the majority-correct event $\{Y > K/2\}$ maps to $\{Y < K/2\}$, so $P_{\mathrm{agg}}(K,q) = 1 - P_{\mathrm{agg}}(K,p)$ is *anti*-symmetric about $1/2$. Thus visible dissent is side-blind: it records that the panel is divided, not which side is correct. The assumption $p > 1/2$ is therefore unnecessary for the closed form itself; it is needed only when $P_{\mathrm{agg}}$ enters, and with it the rate comparison, the $4/5$ constant, and $K^*$.

**Near-agreement points to the correct side.** Conditional on a near-unanimous panel ($Y \le 1$ or $Y \ge K-1$), the side indicated by the panel is correct with probability $p^{K-1}(Kq+p)/\left[p^{K-1}(Kq+p) + q^{K-1}(Kp+q)\right]$. This probability approaches 1 because $(q/p)^{K-1} \to 0$. This is the statement used in the main text.

## Supplementary Note 3 | Dissenting fraction

With $m(K) = \min(Y, K-Y)/K$ and an analysis threshold $r \in (0, \frac{1}{2})$,

$$\lim_{K\to\infty} P\big(m(K) \ge r\big) = \begin{cases} 1, & r < q, \\ \frac{1}{2}, & r = q, \\ 0, & r > q, \end{cases} \qquad \Longrightarrow \qquad m(K) \xrightarrow{p} q.$$

By the SLLN $Y/K \to p$, so $m(K) \to \min(p,q) = q$. At $r = q$ the central limit theorem (CLT) gives limiting probability $\frac{1}{2}$.

## Supplementary Note 4 | Convergence rates, the $p = 4/5$ line, and the $r$-family

Write $E = 1 - P_{\text{agg}}$ and $N = 1 - P_{\text{vis}}$ for the two residuals, with $P_{\text{vis}}$ the fixed-count split of the main text (at least two votes on each side). Both residuals decay exponentially, in the sense of logarithmic limits: $\lim_{K\to\infty} K^{-1}\log E = -R_{\text{agg}}$ and $\lim_{K\to\infty} K^{-1}\log N = -R_{\text{vis}}$ (polynomial prefactors, discussed below, do not affect these limits). By Cramér's theorem,[4] the rate for aggregate reliability and the rate for visible dissent under the relative-threshold definition of Supplementary Note 3 are

$$R_{\text{agg}}(p) = D(\tfrac{1}{2}\,\|\,p) = \tfrac{1}{2}\log\tfrac{1}{4pq}, \qquad R_{\text{vis}}(r,p) = D(1-r\,\|\,p),$$

where $D(a\,\|\,p) = a\log\frac{a}{p} + (1-a)\log\frac{1-a}{q}$. Here $R_{\text{vis}}(r,p)$ is the rate of the $r$-form of visible dissent (at least a fraction $r$ dissenting on each side), defined for $0 < r < q$, the regime in which that event is inevitable (Supplementary Note 3). Its weakest form, $r \to 0$, has rate $R_{\text{vis}} = \log(1/p)$, the same exponential rate as the unanimous outcome $\{Y = K\}$. This limiting rate is exactly that of the main-text fixed-count split, whose dominant near-unanimous edge is $\{Y = K-1\}$. Hence

$$R_{\text{vis}} > R_{\text{agg}} \iff \log\tfrac{1}{p} > \tfrac{1}{2}\log\tfrac{1}{4pq} \iff p < \tfrac{4}{5}.$$

Equivalently, the boundary is $p^2 = 4pq$, that is, $p/q = 4$: the line sits exactly where a single adviser's odds of being correct reach 4 to 1. For the fixed-count split, the limit of $N$ follows directly from the closed form of Supplementary Note 2, whose dominant term is $Kqp^{K-1}$, and the limit of $E$ is Cramér's theorem for the lower tail $\{Y \le K/2\}$. Consequently $\lim_{K\to\infty} K^{-1}\log(E/N) = R_{\text{vis}} - R_{\text{agg}}$, which is positive for every $1/2 < p < 4/5$: there $E/N \to \infty$, so a finite crossing $K^*(p)$ exists and visible dissent eventually remains more likely than a correct majority at every sufficiently large panel size.

For the relative-threshold family, the speed boundary $r^*(p)$ solves $D(1-r^*\,\|\,p) = R_{\text{agg}}(p)$. Since $D(1-r\,\|\,p)$ decreases in $r$, from $\log(1/p)$ at $r = 0$ to 0 at $r = q$, the root exists and is unique exactly when $R_{\text{agg}}(p) < \log(1/p)$, that is, below the line. It stays below $q$ for $p < 4/5$ and collapses to 0 at $p = 4/5$, partitioning the $(p,r)$ plane into three regions (Supplementary Fig. 1).

**Asymptotic versus finite $K$.** The exponents determine only which quantity eventually dominates as $K \to \infty$. The finite-$K$ crossing also depends on the prefactors: the visible-dissent residual carries a factor of order $K$, whereas the majority-error residual carries the usual central-limit factor $K^{-1/2}$.

**Invariance of the 4/5 boundary to fixed residual weighting.** The 4/5 boundary is unchanged by any fixed multiplicative weight on the complementary residuals. For $p < 4/5$ and any fixed $w > 0$, replace $1 - P_{\text{vis}} \le 1 - P_{\text{agg}}$ by $w(1 - P_{\text{vis}}) \le 1 - P_{\text{agg}}$. This shifts the finite crossover but not the asymptotic boundary. Because both residuals decay exponentially, the fixed factor shifts the crossover by roughly $\log w/(R_{\text{vis}} - R_{\text{agg}})$ without changing which residual eventually dominates. Thus fixed residual weighting can move the disclosure reference scale, but not the $p = 4/5$ boundary.

**Any definition of a split.** The bound extends beyond the $r$-family. Under any definition in which a unanimous panel does not count as split, the non-split event contains $\{Y = K\}$, so $1 - P_{\text{vis}} \ge p^K$ and $R_{\text{vis}} \le \log(1/p)$. Since $\log(1/p) > R_{\text{agg}}$ only for $p < 4/5$, no such definition can approach certainty at a faster exponential rate than aggregate reliability at or above $p = 4/5$; the minimal fixed-count split of the main text attains the bound.

**The probability comparison involves disjoint outcomes.** Let $A$ denote visible dissent and $B$ the majority-correct event (with fair tie-breaking, so $P(B) = P_{\text{agg}}$). Every consultation falls into one of four cells (Supplementary Table 1), and the two probabilities share the cell $A \cap B$, in which the panel splits but the majority is correct. Therefore $P_{\text{vis}} - P_{\text{agg}} = P(A \cap B^c) - P(A^c \cap B)$. In words, the probability comparison reduces to two disjoint outcomes: an incorrect-majority split and a correct near-unanimous panel. The shared cell is large under independence (probability 0.89 at $p = 0.7$, $K = 14$), but it cancels exactly. At $p = 0.7$, the incorrect-majority split has probability 0.0624 at both $K = 13$ and $K = 14$, while the correct near-unanimous outcome falls from 0.0637 to 0.0475; this is the crossing $K^*(0.7) = 14$ in Supplementary Note 5. The marginal and disjoint-outcome formulations therefore cross at the same $K^*$.

**Supplementary Table 1. The four outcomes behind the comparison of $P_{\text{vis}}$ with $P_{\text{agg}}$.** $P_{\text{vis}}$ sums the probabilities in the split row, whereas $P_{\text{agg}}$ sums those in the majority-correct column. Tied votes are assigned to either column with equal probability under fair tie-breaking. The shared cell $A \cap B$ cancels in the comparison, leaving the two off-diagonal cells.

| | Majority correct ($B$) | Majority incorrect ($B^c$) |
|---|---|---|
| Split ($A$) | $A \cap B$: counted in both | $A \cap B^c$: counted only in $P_{\text{vis}}$ |
| Near-unanimous ($A^c$) | $A^c \cap B$: counted only in $P_{\text{agg}}$ | $A^c \cap B^c$: counted in neither |

**Heterogeneous panels.** The homogeneous model is the equal-competence special case of a Poisson-binomial panel. For a sequence of independent heterogeneous panels with mean accuracy $\bar{p}_K = K^{-1}\sum_i p_i \to \bar{p} > 1/2$, aggregate reliability still satisfies $P_{\text{agg}} \to 1$,[5] while $Y/K \to \bar{p}$ and hence $m \to 1 - \bar{p}$. The finite and asymptotic crossing behaviour, however, depends on

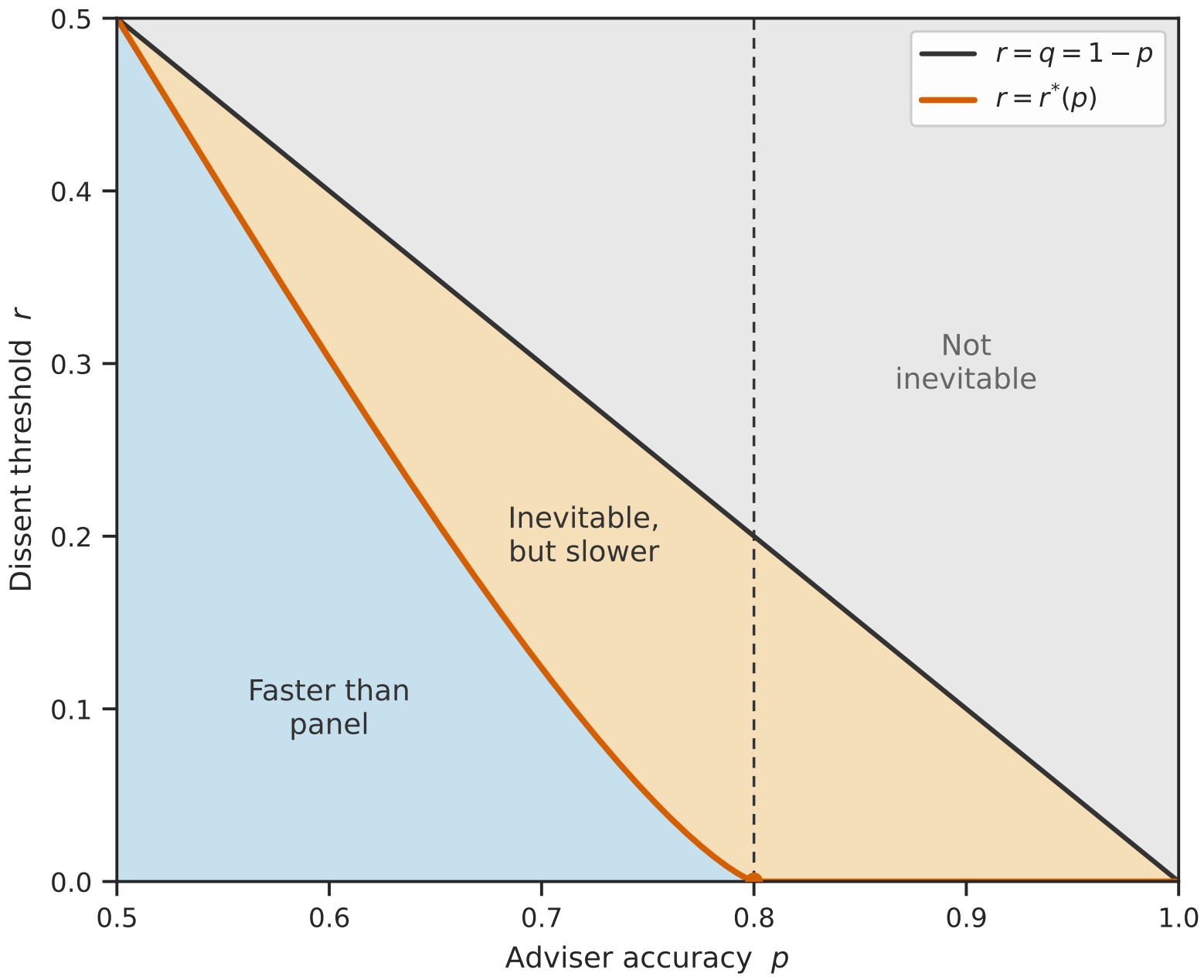


**Supplementary Fig. 1. Phase diagram of visible dissent in $(p, r)$.** The existence boundary $r = q$ and the speed boundary $r = r^*(p)$ partition the plane into *faster than the panel*, *inevitable but slower*, and *not inevitable*. The first of these regions closes at $p = 4/5$. The boundaries are large-deviation rate boundaries (the $K \to \infty$ ordering), with small-$K$ ordering set by the finite crossover $K^*$.

the full accuracy profile $\{p_i\}$, not on the mean alone. For example, a panel with half $p_i = 0.999$ and half $p_i = 0.55$ has mean accuracy $\bar{p} \approx 0.77$, below $4/5$, yet a correct majority remains more likely than a split at every panel size checked numerically up to $K = 400$. Mean accuracy alone therefore does not locate the crossing. When the per-adviser accuracies $\{p_1, \ldots, p_K\}$ are available, compute $P_{\text{vis}}$ and $P_{\text{agg}}$ directly from the Poisson-binomial law and compare them for that panel ($K \geq 4$, fair tie-breaking). The $4/5$ line and $K^*(p)$ are exact summaries only for the equal-competence case; for independent heterogeneous panels, compute both probabilities from the full accuracy profile rather than applying $4/5$ to the mean accuracy.

## Supplementary Note 5 | Disclosure reference scale $K^*$

Define

$$K^*(p) = \min\{K : P_{\text{vis}}(K, p) \geq P_{\text{agg}}(K, p)\},$$

the first panel size at which visible dissent becomes at least as likely as a correct majority. The values in the main text use fair tie-breaking: for example, $K^*(0.7) = 14$ (Fig. 3b of the main text), whereas a strict-majority convention gives $K^*(0.7) = 10$. For every $p < 4/5$, Supplementary Note 4 guarantees that such a finite crossing exists. Before $K^*$, $P_{\text{agg}} > P_{\text{vis}}$ by definition.

The two parity classes cross at different sizes, so immediately above $K^*$ the ordering can depend on parity. At $p = 0.735$, for example, $P_{\text{vis}} \geq P_{\text{agg}}$ holds at $K = 22$ (0.98978 versus 0.98976) and at $K = 24$ (0.99404 versus 0.99230) but not at $K = 23$ (0.99219 versus 0.99230): the even class has crossed at $K = 22$, the odd class only at $K = 25$. Statements about a given panel size must therefore compare the two probabilities at that size; $K^*$ marks the first crossing only. For every $p < 4/5$, visible dissent eventually remains more likely than a correct majority (Supplementary Note 4).

Near the line, both probabilities can round to 1 in double precision, producing spurious crossings; the accompanying scripts therefore locate the crossover by comparing the complementary tails in log space.

The rate comparison itself does not depend on how ties are broken; only this finite $K^*$ does. Under the strict convention a tie counts as a failure, so $P_{\text{agg}}$ is lower at even $K$, visible dissent overtakes sooner, and the strict $K^*$ is never larger than the fair one.

Near $p = 4/5$, $K^*(p)$ grows rapidly, so small errors in the estimate of $p$ can produce large uncertainty in the estimated reference scale (Fig. 3b of the main text).

**The regime $p \geq 4/5$, including the boundary.** That reliability keeps its lead at every finite $K$ on and above the line (Fig. 2a of

the main text) cannot be read from the figure alone. For $p \geq 4/5$ one has $\sqrt{4pq} \leq p$, so

$$1 - P_{\text{vis}}(K) \geq p^{K-1}(Kq + p) = (1 + Kq/p)\, p^K \geq (1 + Kq/p)\,(4pq)^{K/2},$$

while the Chernoff–Hoeffding bound gives $1 - P_{\text{agg}}(K) \leq e^{-KD(\frac{1}{2}\,\|\,p)} = (4pq)^{K/2}$.[4] Hence $P_{\text{vis}}(K) < P_{\text{agg}}(K)$ for every $K \geq 4$ and every $p \geq 4/5$, so no finite $K^*$ exists on or above the line. On the line itself the lead is an ordering, not a usable margin: both residuals vanish exponentially, and at $p = 4/5$, $K = 100$ the gap $P_{\text{agg}} - P_{\text{vis}}$ is about $5 \times 10^{-9}$.

## Supplementary Note 6 | $P_{\text{vis}}$ as a null model

The main text uses $P_{\text{vis}}(K, p)$ as an independence baseline. In the one-factor Gaussian copula settings illustrated in Supplementary Fig. 2, positive correlation reduces visible disagreement, producing a "dissent deficit" relative to that baseline. Positive pairwise correlation alone does not imply a lower split rate. For example, a four-adviser exchangeable panel with $\Pr(Y = 4) = 0.5$, $\Pr(Y = 2) = 0.4$, $\Pr(Y = 0) = 0.1$ has individual accuracy 0.7 and pairwise correlation $\approx 0.37$, yet $P_{\text{vis}} = 0.40$ against 0.26 under independence. Nor does positive pairwise correlation necessarily reduce majority accuracy. For example, a three-adviser exchangeable panel with $\Pr(Y = 3) = 0.5$, $\Pr(Y = 2) = 0.3$, $\Pr(Y = 0) = 0.2$ also has individual accuracy 0.7 and positive pairwise correlation ($\approx 0.52$), yet its majority accuracy is 0.80, compared with 0.784 under independence. Accordingly, a deviation of the adjusted split rate from the baseline in either direction, beyond the uncertainty of the accuracy estimate and sampling variation, provides evidence of residual dependence. Correlated extensions of the theorem[6,7] concern aggregate reliability rather than the split rate. In the simulated Gaussian copula settings, the deficit provides an operational diagnostic. For a homogeneous reference panel, the deficit can be summarized by an effective panel size $K_{\text{eff}} < K$. When per-adviser accuracies are available, the Poisson-binomial comparison in Supplementary Note 4 should be used directly, with $K_{\text{eff}}$ only as a secondary composite summary. Empirical validation of this diagnostic on real panels is left to future work. Concurrent evidence suggests the deficit can be severe. An audit of an LLM evaluation panel found nine frontier judges supplying only about two effective independent votes.[8] It estimated that effective size from pairwise error correlations rather than from the split rate.

At a fixed mean accuracy, adviser heterogeneity can either raise or lower the split rate. Unequal competences reduce the probability of full unanimity ($\prod_i p_i \leq \bar{p}^K$ by the arithmetic–geometric mean inequality) but can increase the probability of a one-dissenter outcome. At small $K$, for example, one weak adviser can act as a consistent lone dissenter, keeping the split rate below the equal-competence baseline. When per-adviser accuracies are available, the Poisson-binomial baseline accounts for this heterogeneity exactly. Otherwise, a dissent deficit may reflect dependence, heterogeneity, or both, so $K_{\text{eff}}$ is a composite diagnostic rather than a measure of independence alone. Supplementary Fig. 2 illustrates two effects in synthetic correlated panels: the dissent deficit (a) and the inflated all-wrong unanimity rate (b). In panel (a), at $K^*(0.7) = 14$, the correlated split rates correspond to $K_{\text{eff}} \approx 9.4$ for $\rho$=0.2 and $K_{\text{eff}} \approx 6.6$ for $\rho$=0.5.

**A diagnostic procedure for the disclosure layer.** For a homogeneous reference panel, or when only a pooled accuracy estimate is available, the steps below turn the dissent deficit into a diagnostic input. Let $K$ be the panel size and let $\hat{p}$ be an accuracy estimate obtained from held-out labelled items or, more coarsely, from a reported benchmark:

(i) Compute the independence baseline $P_{\text{vis}}(K, \hat{p})$.

(ii) Measure the observed split rate $\hat{s}$, the fraction of items on which the panel visibly divides (at least two verdicts on each side) across a sample.

(iii) If $\hat{s} < P_{\text{vis}}(K, \hat{p})$ after accounting for item difficulty and adviser heterogeneity, and the gap exceeds the uncertainty of the accuracy estimate and sampling variation, treat it as evidence of residual dependence. For a homogeneous reference, summarize its size by $K_{\text{eff}} \leq K$, defined through $P_{\text{vis}}(K_{\text{eff}}, \hat{p}) = \hat{s}$; non-integer values use the continuous extension of the closed form.

Item difficulty and unequal adviser accuracies can both mimic a dissent deficit. Where possible, stratify items by difficulty and use the Poisson-binomial baseline when per-adviser accuracies are available. The reason is the nonlinearity in $p$: averaging accuracy first does not reproduce either $\mathbb{E}_i[P_{\text{vis}}(K, p_i)]$ or $\mathbb{E}_i[P_{\text{agg}}(K, p_i)]$, so $K^*(\bar{p})$ has the same limitation. After these corrections, $K_{\text{eff}}$ summarizes whatever deficit remains. Difficulty stratification itself requires per-item difficulty estimates, which usually require labels. Whether the way the split rate changes with $K$ can separate dependence from difficulty mixing without labels remains an open question.

**Mean accuracy does not determine the ordering.** As a numerical illustration, take $K = 14$ and two item populations with the same mean accuracy $\bar{p} = 0.7$: a homogeneous population with $p = 0.7$ on every item, and an equal mixture of items with $p = 0.51$ and $p = 0.89$. Within each item, advisers remain independent with equal accuracy, and the probabilities are averaged over items (Supplementary Table 2). In the homogeneous population, $P_{\text{vis}}$ exceeds $P_{\text{agg}}$ at $K = 14$, which is $K^*(0.7)$; in the mixture, the order at the same $K$ is reversed. Mean accuracy alone therefore determines neither the ordering at a given panel size nor the location of a crossing, which is why the procedure above stratifies by difficulty before comparing with the baseline.

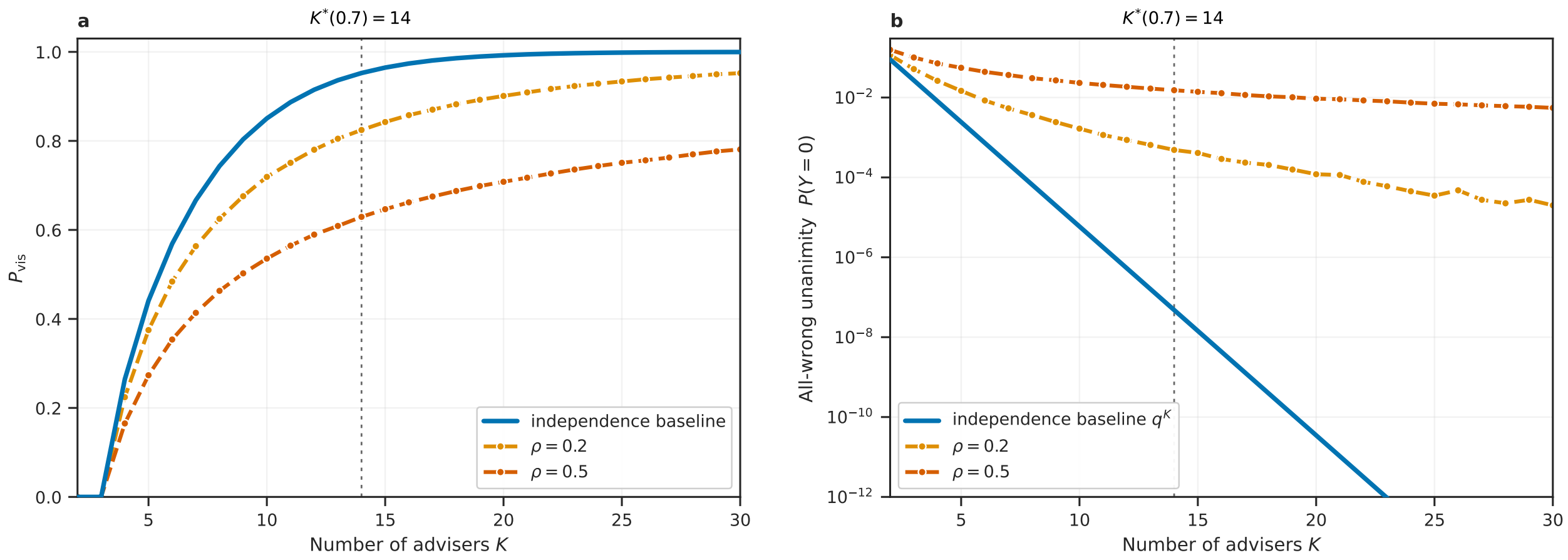


**Supplementary Fig. 2. The independence baseline as a null model (illustrative).** Synthetic correlated panels (Gaussian copula, $\rho = 0.2, 0.5$, dashed) are compared with the closed-form independence baselines (solid) at $p = 0.7$; the vertical dotted line marks the disclosure reference scale $K^*(0.7) = 14$. **(a)** Correlation pulls the split rate below the baseline $P_{\text{vis}}(K, p)$; all curves are zero for $K < 4$, where a two-per-side split is impossible. At $K^*$ the baseline predicts $P_{\text{vis}} \approx 0.95$, whereas the correlated panels split on about 82.5% ($\rho$=0.2) and 63% ($\rho$=0.5) of items. **(b)** The same correlation inflates the all-wrong unanimity rate $P(Y=0)$ (log scale): at $K^*$, $\approx 5 \times 10^{-4}$ ($\rho$=0.2) and $\approx 1.5 \times 10^{-2}$ ($\rho$=0.5) against the independence baseline $q^K \approx 5 \times 10^{-8}$, factors of $10^4$ and $3 \times 10^5$ above it. Simulation design: one-factor Gaussian copula, $Z_i = \sqrt{\rho}\, Z_0 + \sqrt{1-\rho}\, \varepsilon_i$ with $Z_0, \varepsilon_1, \ldots, \varepsilon_K$ independent standard normal variables, adviser $i$ correct if and only if $Z_i < \Phi^{-1}(p)$, so that $\rho$ is the correlation of the latent normal variables and each adviser is correct with probability $p$; 400,000 simulated items per $(K, \rho)$ cell, $K = 4, \ldots, 30$, seed 7 (`scripts/run_figS2_nullmodel.py`, then `scripts/figures/figS2_nullmodel.py`). Synthetic simulation, not real-panel data.

**Supplementary Table 2. Same mean accuracy, opposite ordering at $K = 14$.** Item-averaged probabilities under independence and equal accuracy within each item; exact values rounded to four decimals.

| Item population | $\mathbb{E}_i[P_{\text{agg}}(14, p_i)]$ | $\mathbb{E}_i[P_{\text{vis}}(14, p_i)]$ |
|---|---|---|
| $p = 0.7$ on every item | 0.9376 | 0.9525 |
| $p = 0.51$ items only (mixture component) | 0.5293 | 0.9981 |
| $p = 0.89$ items only (mixture component) | 0.9998 | 0.4658 |
| Equal mixture of 0.51 and 0.89 | 0.7646 | 0.7320 |

In these simulations, correlation also raises the probability of an all-wrong panel above the independence baseline $q^K$ (Supplementary Fig. 2b), so the over-trust caution of the main text applies alongside this procedure. At the correlations of Supplementary Fig. 2, that inflation means one unanimously wrong item in roughly 2,000 ($\rho$=0.2) and one in roughly 66 ($\rho$=0.5), each presented as maximal apparent certainty with nothing in the verdicts to flag it, versus one in twenty million under independence.

## Supplementary Note 7 | Three user-side quantities and the monotone-response hypothesis

All three literatures read the same vote split $Y$; the mapping is explicit, and for predictive multiplicity it is exact. **Predictive multiplicity**, translated into the panel setting, is the rate at which the panel disagrees. If "disagreement" means any non-unanimous outcome, so that one dissenter is enough, its probability is $P_{\text{any}} = 1 - p^K - q^K$; for $K = 2$ this reduces to the pairwise rate $2pq$ below. If disagreement is defined by the minimal split in Supplementary Note 2, the probability is $P_{\text{vis}}(K, p)$. The disclosure framework of the main text is formulated in terms of this second definition; Fig. 4a of the main text plots the first. The two events differ only in the one-dissenter cases $\{Y = 1\} \cup \{Y = K - 1\}$, so their complements have the same asymptotic rate $\log(1/p)$; the 4/5 rate boundary therefore holds under either definition. The connection to the predictive-multiplicity literature concerns the observed disagreement rate, not how the competing models were generated. In that literature the models are typically drawn from a level set fitted to one dataset, so they need not be independent or sampled as in the i.i.d. model here. **Reconciliation load** is linked here to the expected dissenting fraction $\mathbb{E}[m(K)]$, which we treat as a candidate stimulus for that

load. The expected fraction approaches $q$ as the panel grows.

**Reliance miscalibration** has two directions. For under-reliance, the stimulus is visible dissent, which occurs with probability $P_{\text{vis}}(K,p)$. Because most splits occur on items whose majority is actually correct, repeated exposure to dissent can encourage undue doubt. For over-trust, the observable cue is apparent unanimity $U = p^K + q^K$; its all-wrong component $q^K$ is the probability of exposure to a unanimously wrong panel, which looks the same as a correct one. Following unanimity is well founded under independence; over-trust means reliance beyond what the evidence warrants, as when correlation inflates the all-wrong component but users treat the signals as independent. Under independence, the all-wrong probability $q^K$ decreases exponentially with panel size, and its ratio to the all-correct probability is $(q/p)^K \to 0$; correlation can inflate it substantially (Supplementary Note 6).

**Pairwise disagreement.** Two independent binary advisers disagree with probability $2pq$, one correct and one wrong in either order, so within the binary model an observed pairwise disagreement rate $\delta$ would imply $p = \frac{1}{2}\left(1+\sqrt{1-2\delta}\right)$. This inversion is not applied to empirical rates in the main text. For multiple-choice answers, $\delta = 2pq + q^2\,\Pr(\text{different wrong answers} \mid \text{both wrong})$, so the binary formula understates $p$; correlation, unequal accuracies and difficulty mixtures shift $\delta$ further. The identity serves only to translate the model's pairwise rate into the vocabulary of predictive multiplicity.

**The monotone-response hypothesis (precise form).** Predictive multiplicity is directly identified with an observable disagreement probability, but reconciliation load and reliance are behavioural responses to split-derived stimuli. We therefore assume only monotone relationships: expected reconciliation load does not decrease as $\mathbb{E}[m(K)]$ increases, and expected under-reliance on majority-correct items does not decrease as exposure to visible dissent $P_{\text{vis}}(K,p)$ increases. Over-trust is not part of the hypothesis: for it, the model supplies the probability $q^K$ of exposure to unanimously wrong advice, and whether reliance on apparent unanimity exceeds what the evidence warrants is an empirical question. These are behavioural hypotheses, not results of the mathematical model.

**The prediction it yields.** Because $P_{\text{vis}}$ and $\mathbb{E}[m]$ are explicit functions of $(K,p)$, the hypothesis fixes both the direction and the $(K,p)$-ordering of the behavioural effects. Holding $p$ fixed, increasing $K$ raises $P_{\text{vis}}$ and does not decrease $\mathbb{E}[m]$, so the modelled stimulus for under-reliance increases while that for reconciliation load does not decrease. Whether the behavioural responses themselves become pronounced near $K^*(p)$ is an empirical question. Crossing the $4/5$ line changes the relative ordering of $P_{\text{vis}}$ and $P_{\text{agg}}$ at large $K$ (a finite $K^*$ exists only below it), not the direction in which each stimulus moves with $K$.

**Testing the hypothesis.** A test exposes users to panels of varying $(K,p)$ and measures reliance and reported load against the predicted $P_{\text{vis}}$ and $\mathbb{E}[m]$; evidence that the expected response decreases as its corresponding stimulus increases, after accounting for other relevant factors, would contradict the hypothesis, while a $(K,p)$-independent response would instead show that behaviour is insensitive to the modelled stimuli. Known moderators of reliance, such as explanation quality, task domain, and user expertise, enter such a test as covariates; the hypothesis concerns the ordering induced by the stimuli, not the absence of other influences.

**Confidence conditional on the observed vote.** When presentation policies are compared as in Box 1 of the main text, specify how confidence is estimated, distinguishing model-based aggregate accuracy from confidence conditional on the observed vote. Let $d = |2Y-K|$ be the absolute vote margin. Given $d \geq 1$ and a fixed $p$, the majority is correct with probability $1/(1+(q/p)^d)$, which depends on the observed votes only through the absolute margin; a tie ($d=0$) is resolved by fair tie-breaking and is correct with probability $1/2$. At $p=0.7$ and $K=14$ the unconditional $P_{\text{agg}}$ is 0.938, but 0.845 given an 8–6 split. In behavioural tests, declining to follow a majority that turned out to be correct is not by itself inappropriate reliance; appropriateness is judged against the information available at the time, including the margin.



\end{document}